\documentclass{sig-alternate}

\PassOptionsToPackage{table}{xcolor}
\usepackage{graphicx}
\usepackage{stmaryrd}
\usepackage{tikz}
\usetikzlibrary{shapes,arrows,shapes.geometric,fit,matrix,backgrounds}
\PassOptionsToPackage{table}{xcolor}
\usepackage{graphicx}
\usepackage{caption}
\usepackage{subcaption}
\usepackage{framed}
\usepackage{fancybox}
\usepackage{colortbl}
\usepackage{color}
\usepackage{hyperref}
\usepackage{pgfpages}
\usepackage{amsmath,mathtools,amssymb}
\usepackage{balance}
\usepackage{lineno}
\newtheorem{myex}{Example}
\newtheorem{myprob}{Challenge}
\newtheorem{myremark}{Remark}

\begin{document}
%
\conferenceinfo{XXX}{XXX}

\title{\huge{$\boldsymbol{\Upsilon}$}\ttlfnt{-DB: A system for data-driven hypothesis\\ management and analytics}}

\numberofauthors{3} 
%
\author{
%
\alignauthor
Bernardo Gon\c{c}alves\\
       \affaddr{LNCC -- National Laboratory}\\
       \affaddr{for Scientific Computing}\\
       \affaddr{Petr\'opolis, Brazil}\\
       \email{bgonc@lncc.br}
\alignauthor Frederico C. Silva\\
       \affaddr{LNCC -- National Laboratory}\\
       \affaddr{for Scientific Computing}\\
       \affaddr{Petr\'opolis, Brazil}\\
       \email{fredcs@lncc.br}\\
\alignauthor Fabio Porto\\
       \affaddr{LNCC -- National Laboratory}\\
       \affaddr{for Scientific Computing}\\
       \affaddr{Petr\'opolis, Brazil}\\
       \email{fporto@lncc.br}
}

\maketitle

\begin{abstract}
The vision of $\Upsilon$-DB introduces deterministic scientific hypotheses as a kind of uncertain and probabilistic data, and opens some key technical challenges for enabling data-driven hypothesis management and analytics. The $\Upsilon$-DB system addresses those challenges throughout a design-by-synthesis pipeline that defines its architecture. It processes hypotheses from their XML-based extraction to encoding as uncertain and probabilistic U-relational data, and eventually to their conditioning in the presence of observations.
In this demo we present a first prototype of the $\Upsilon$-DB system. We showcase its core innovative features by means of use case scenarios in computational science in which the hypotheses are extracted from a model repository on the web and evaluated (rated/ranked) as probabilistic data.
\end{abstract}


\vspace{-6pt}

\vspace{-3pt}
\section{Introduction}
\noindent
As part of the paradigm shift that makes science ever more data-driven, deterministic scientific hypotheses shall be seen as principles or ideas, which are mathematically expressed and then implemented in a program that is run to give their \emph{decisive} form of data. For a description of the research vision of hypothesis management and its significance, we refer the reader to \cite{goncalves2014}. In this demo we present a first prototype of the $\Upsilon$-DB system. We introduce its key research challenges and showcase its main innovative features by means of use cases in computational science.

\textbf{Target applications.} 
Our framework is geared towards hypothesis management applications. Examples of structured deterministic hypotheses include tentative mathematical models in physics, engineering and economical sciences, or conjectured boolean networks in biology and social sciences. These are important reasoning devices, as they are solved to generate predictive data for decision making in both science and business. But the complexity and scale of modern scientific problems require proper data management tools for the predicted data to be analyzed more effectively.

$\!\!$\textbf{Probabilistic DBs.} 
Probabilistic databases $\!$(p-DBs) qualify as such tool, as they have evolved into mature technology in the last decade 
\cite{suciu2011}. One of the state-of-the-art probabilistic data models is the U-relational representation system with its probabilistic world-set algebra (p-WSA) implemented in \textsf{MayBMS} $\!$\cite{koch2009}. $\!$That is an elegant extension of the relational model we have adopted for the management of uncertain and probabilistic data. 
It is a core feature of \mbox{$\Upsilon$-DB} the capability to extract a hypothesis specification and encode it into a U-relational DB seamlessly, ensuring consistency and quality w.r.t.$\!$ the given hypothesis structure, while alleviating the user from the burden of p-DB design. 
In short, the system flattens tentative deterministic models into U-relations for data-driven management and analytics. 

\begin{figure}[t]
\tikzstyle{rect1}=[rectangle,
                                    thick,
                                    minimum size=15pt,
                                    draw=black]
\tikzstyle{rect2}=[rectangle,
                                    thick,
                                    minimum size=15pt,
                                    fill=black!20,
                                    draw=black]
\tikzstyle{rect3}=[rectangle,
                                    rounded corners=3pt,
                                    minimum size=60pt,
                                    minimum width=80pt,
                                    draw=black]
\tikzstyle{box}=[rectangle,
                                    fill=none,
                                    draw=none]
\tikzstyle{cyl1}=[cylinder,
                                    thick,
                                    minimum size=23pt,
                                    inner sep=0pt,
                                    fill=none,
                                    draw=black]
\tikzstyle{cyl2}=[cylinder,
                                    thick,
                                    fill=black!20,
                                    minimum size=23pt,
                                    inner sep=0pt,
                                    draw=black]
\tikzstyle{edge} = [draw,thick,->,bend left]
\begin{tikzpicture}[scale=0.85]
    \node[rect3] (back) at (0,3) {};
    \node[rect1] (s) at (0,3.7) {$\mathcal{S}_k$};
    \node[rect2] (d1) at (-1.1,2.25) {$\mathcal{D}_k^1$};
    \node[rect2] (d2) at (-0.2,2.25) {$\mathcal{D}_k^2$};
    \node[box] (dot) at (0.45,2.25) {...};
    \node[rect2] (dn) at (1.1,2.25) {$\mathcal{D}_k^p$};
    \node[cyl1,rotate=90] (h) at (4,3) {\rotatebox[origin=c]{-90}{h}};
    \node[box] (hlabel) at (4,2.1) {$\bigcup_{k=1}^{z}\bigcup_{i=1}^{n}H_k^i$};
    \node[cyl2,rotate=90] (y) at (7.1,3) {\rotatebox[origin=c]{-90}{y}};
    \node[box] (ylabel) at (7.1,2.1) {$\bigcup_{k=1}^{z}\bigcup_{j=1}^{m}Y_k^j$};
    \node[box] (cond) at (7.1,4) {\huge\rotatebox[origin=c]{180}{$\circlearrowleft$}};
    \draw[-] (s) to (d1);
    \draw[-] (s) to (d2);
    \draw[-] (s) to (dn);
    \draw[->] (back) to (h);
    \node[box] (etl) at (2.6,3.3) {\textsf{ETL}};    
    \draw[->] (h) to (y);
    \node[box] (etl) at (5.55,3.3) {\textsf{U-intro}};
    \node[box] (etl) at (7.1,4.5) {\textsf{conditioning}};
\end{tikzpicture}
\vspace{-12pt}
\caption{Design-by-synthesis pipeline for processing hypotheses as uncertain and probabilistic data.}
\label{fig:pipeline}
\vspace{-7pt}
\end{figure}
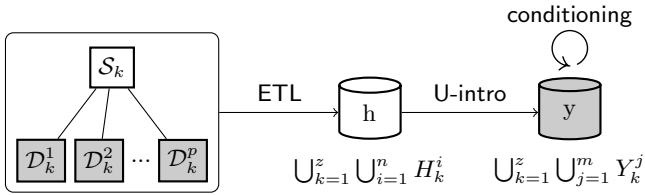

It comprises some key technical challenges that are dealt with elsewhere \cite{goncalves2015a}. The demonstrated system addresses them throughout a design-by-synthesis pipeline that defines its architecture, see Fig. \ref{fig:pipeline}. It processes hypotheses from their XML-based extraction to encoding as uncertain and probabilistic U-relational data, and eventually to their conditioning in the presence of observations. 

We introduce next (\S\ref{sec:pipeline}) the pipeline in some detail by emphasizing its main technical challenges. Then in \S\ref{sec:related-work} we discuss related work and in \S\ref{sec:applicability} the system applicability (assumptions, scope, etc). Finally, in \S\ref{sec:demo} we describe the demo.

\section{Design-By-Synthesis Pipeline}\label{sec:pipeline}
We refer to Example \ref{ex:population} for an accessible and straightforward illustration of the technical challenges addressed by the demonstrated system throughout its pipeline architecture.

\begin{myex}
We explore three different theoretical models in population dynamics with applications in Ecology, Epidemics, Economics, etc: (\ref{eq:malthus}) Malthus' model, (\ref{eq:logistic})$\!$ the logistic equation and (\ref{eq:lotka-volterra}) the Lotka-$\!$Volterra model. In the demonstration, such equations are extracted from \textsf{MathML}-compliant XML files (cf. \S\ref{sec:demo}). For now, consider that the ordinary differential equation notation `$\dot{x}$' is read `variable $x$ is a function of time $t$ given initial condition $x_0$.$\!$' These three hypotheses are considered as competing explanations for two phenomena, viz., (1) the nationwide US population from 1790 to 1990, and (2) the Lynx population in Hudson's Bay, Canada, from 1900 to 1920. $\Box$
\label{ex:population}
\end{myex}
\begin{eqnarray}
\dot{x}=b\,x
\label{eq:malthus}
\end{eqnarray}
\vspace{-18pt}
\begin{eqnarray}
\dot{x}=b\,(1-x/K)\,x
\label{eq:logistic}
\end{eqnarray}
\vspace{-18pt}
\begin{eqnarray}
\left\{ 
  \begin{array}{lll}
\dot{x} &=& x\,(b - p\,y)\\
\dot{y} &=& y\,(r\,x - d)
\end{array} \right.
\label{eq:lotka-volterra}
\end{eqnarray}

\subsection{Hypothesis Encoding}
\noindent
\textbf{Structural equations.} 
In the form of mathematical equations, hypotheses symmetrically relate aspects of the studied phenomenon. For computing predictions, however, deterministic hypotheses are used asymmetrically as \emph{functions} \cite{simon1966}. They take a given valuation over input variables (parameters) to produce values of output variables (the predictions). Such an asymmetry establishes functional dependencies that can be discovered in the predictive data. 
 
Formally, given a system $\mathcal S(\mathcal E, \mathcal V)$ of mathematical equations with a set of variables appearing in them, the AI literature has seen the introduction of an asymmetrical, functional relation among variables that establishes a \emph{causal ordering} \cite{simon1953,druzdzel2008}. The algorithmic inference of the causal ordering embedded in a deterministic system $\mathcal S$ is a non-trivial challenge for transforming it into a \emph{structural equation model} (SEM), viz., a total mapping $\varphi_t\!:\, \mathcal E \to \mathcal V$ from the set $\mathcal E$ of equations to the set $\mathcal V$ of variables. That is a key technical challenge we have addressed in the $\Upsilon$-DB system to enable the encoding of a hypotheses as data \cite{goncalves2015a}. It lies at the core of an algorithm we have designed to rigorously transform the hypothesis structure (its equations given in \textsf{MathML} format)\footnote{\textsf{MathML} is W3C's standard for encoding mathematical information (\url{http://www.w3.org/Math/}).}  into a set of fd's.

\begin{myprob} (\textbf{Hypothesis encoding}).\label{prob:h-encode}
Given a deterministic hypothesis $k$ with its structure $\mathcal S_k$, extract a total causal mapping $\varphi_t$ over $\mathcal S_k$ and encode $\varphi_t$ into an fd set $\Sigma_k$.
\vspace{-12pt}
\end{myprob}

A validity condition for a given mathematical system to be grounded into target phenomena such that falsifiable predictions can be derived from it is that $|\mathcal E| \!=\! |\mathcal V|$ \cite{goncalves2015a}. In other words,
in addition to the theoretical equations, subsidiary equations (e.g., $x_0\!=\!200,\, b\!=\!10$), must also be provided such that we know the hypothesis (theory) can be grounded into some target phenomenon by means of a simulation trial. 

\setcounter{myex}{0}
\begin{myex}$\!$(continued).$\!$ 
In this example we are given structures $\mathcal S_k(\mathcal E_k, \mathcal V_k)$ for $k=1..3$ as follows.
\begin{itemize}
\item $\mathcal E_1 \!=\! \{\, f_1(t),\; f_2(x_0),\; f_3(b),\; f_4(x, t, x_0, b) \,\}$;
\item $\mathcal E_2 \!=\! \{ f_1(t),\, f_2(x_0),\, f_3(K),\, f_4(b),\, f_5(x, t, x_0, K, b) \}$;
\item $\mathcal E_3 \!=\! \{\, f_1(t),\; f_2(x_0),\; f_3(b),\; f_4(p),\; f_5(y_0),\; f_6(d),$\vspace{2pt}\\ 
			$f_7(r),\; f_8(x, t, x_0, b, p, y),\; f_9(y, t, y_0, d, r, x) \,\}$.  $\Box$
\end{itemize}
\end{myex}

For instance, the structure of hypothesis $H_3$ given above has $|\mathcal E_3|\!=\!|\mathcal V_3|\!=\!9$. The $\Upsilon$-DB system then accepts it as valid and encodes it into fd set $\Sigma_3$ as shown in Fig. $\!$\ref{fig:population} (left). Note, in all fd sets $\Sigma_k$ with $k=1..3$ that $\phi$ and $\upsilon$ stand (resp.) for the id's of phenomena and hypotheses. This is a core abstraction of $\Upsilon$-DB to capture the data-level semantics of mathematical equations and encode it into fd's \cite{goncalves2014}.

The fd set $\Sigma_k$ of hypothesis $k$ is basic input to schema synthesis at two stages of the pipeline, viz., the \textsf{ETL} and \textsf{U-intro} stages (Fig. \ref{fig:pipeline}) we introduce next.

\begin{figure}[t]\scriptsize
   \centering
\begingroup\setlength{\fboxsep}{3pt}
\colorbox{blue!4}{%
   \begin{tabular}{c|c|p{0.48\linewidth}}
  \textsf{PHENOMENON} & $\phi$ & \textsf{Description}\\
      \hline    
   & $1$ & US population from 1790 to 1990.\\
   & $2$ & Lynx population in Hudson's Bay, Canada, from 1900 to 1920.\\
   \end{tabular}
}\endgroup
\vspace{1pt}
\begingroup\setlength{\fboxsep}{5pt}
\colorbox{blue!4}{%
   \begin{tabular}{c|c|l}
  \textsf{HYPOTHESIS} & $\upsilon$ & \textsf{Name}\\
      \hline    
   & $1$ & Malthus' model\\
   & $2$ & Logistic equation\\
   & $3$ & Lotka-Volterra model\\
   \end{tabular}
}\endgroup
\vspace{-3pt}
\caption{Descriptive (textual) data of Example \ref{ex:population}.}
\label{fig:research}
\vspace{-7pt}
\end{figure}

\begin{figure*}[t]
\footnotesize
\begin{framed}
\vspace{-9pt}
\begin{subfigure}{0.125\columnwidth}
\begin{eqnarray*}
\!\Sigma_1 = \{\;\; \phi &\!\to\!& x_0,\\ 
\phi &\!\to\!& b,\\ 
\!\!x_0\,b\,t\,\upsilon &\!\to\!& x \,\,\}.
\end{eqnarray*}
\begin{eqnarray*}
\Sigma_2 = \{\;\; \phi &\!\to\!& x_0,\\ 
\phi &\!\to\!& K,\\ 
\phi &\!\to\!& b,\\ 
x_0\,K\,b\,t\,\upsilon &\!\to\!& x \;\,\}.
\end{eqnarray*}
\end{subfigure}
\hspace{15pt}
\begin{subfigure}{0.2\columnwidth}
\begin{eqnarray*}
\Sigma_3 = \{\;\;\; \phi &\!\to\!& x_0,\\ 
\phi &\!\to\!& b,\\
\phi &\!\to\!& p,\\
\phi &\!\to\!& y_0,\\ 
\phi &\!\to\!& d,\\ 
\phi &\!\to\!& r,\\
x_0\,b\,p\,t\,\upsilon\,y &\!\to\!& x,\\ 
y_0\,d\,r\,t\,\upsilon\,x &\!\to\!& y \;\,\}.
\end{eqnarray*}
\end{subfigure}
\begin{subfigure}{1.38\columnwidth}
\scriptsize
\begingroup\setlength{\fboxsep}{3pt}
\colorbox{blue!7}{%
   \begin{tabular}{c|>{\columncolor[gray]{0.92}}c||c|c|c|c|c|c|c}
  $\!\!$\textsf{H}$_3^1\!\!$ & $\!\!$\textsf{tid}$\!\!$ & $\!\phi\!$ & $\!x_0\!$ & $b$ & $p$ & $\!y_0\!$ & $d$ & $r$\\
      \hline
   & $\!\!1\!\!$ & $\!2\!$ & $\!30\!$ & $\!\!.5\!\!$ & $\!\!.020\!\!$ & $\!\!4\!\!$ & $\!\!.75\!\!$ & $\!\!.020\!\!$\\
   & $\!\!2\!\!$ & $\!2\!$ & $\!30\!$ & $\!\!.5\!\!$ & $\!\!.018\!\!$ & $\!\!4\!\!$ & $\!\!.75\!\!$ & $\!\!.023\!\!$\\
   & $\!\!3\!\!$ & $\!2\!$ & $\!30\!$ & $\!\!.4\!\!$ & $\!\!.020\!\!$ & $\!\!4\!\!$ & $\!\!.8\!\!$ & $\!\!.020\!\!$\\
   & $\!\!4\!\!$ & $\!2\!$ & $\!30\!$ & $\!\!.4\!\!$ & $\!\!.018\!\!$ & $\!\!4\!\!$ & $\!\!.8\!\!$ & $\!\!.023\!\!$\\
   & $\!\!5\!\!$ & $\!2\!$ & $\!30\!$ & $\!\!.397\!\!$ & $\!\!.020\!\!$ & $\!\!4\!\!$ & $\!\!.786\!\!$ & $\!\!.020\!\!$\\
   & $\!\!6\!\!$ & $\!2\!$ & $\!30\!$ & $\!\!.397\!\!$ & $\!\!.018\!\!$ & $\!\!4\!\!$ & $\!\!.786\!\!$ & $\!\!.023\!\!$\\
   \end{tabular}
}\endgroup
$\;\bowtie\!\!$
\begingroup\setlength{\fboxsep}{3pt}
\colorbox{blue!7}{%
   \begin{tabular}{c|>{\columncolor[gray]{0.92}}c||c|c|c|c|c}
  $\!\!$\textsf{H}$_3^2\!\!$ & $\!\!\textsf{tid}\!\!$ & $\!\phi\!$ & $\!\upsilon\!$ & $\! t \!$ & $y$ & $x$\\
      \hline    
   & $\!\!1\!\!$ & $\!2\!$ & $\!3\!$ & $\!\!1900\!\!$ & $\!\!4\!\!$ & $\!\!30\!\!$\\
   & $\!\!1\!\!$ & $\!2\!$ & $\!3\!$ & $\!..\!$ & .. & ..\\
   \cline{2-7}
   & $\!\!..\!\!$ & $\!2\!$ & $\!3\!$ & $\!..\!$ & .. & ..\\
   \cline{2-7}
   & $\!\!6\!\!$ & $\!2\!$ & $\!3\!$ & $\!\!1900\!\!$ & $\!\!4\!\!$ & $30$\\
   & $\!\!6\!\!$ & $\!2\!$ & $\!3\!$ & $\!\!1901\!\!$ & $\!\!4.12\!\!$ & $\!\!41.5\!\!$\\
   & $\!\!6\!\!$ & $\!2\!$ & $\!3\!$ & $\!\!1902\!\!$ & $\!\!5.78\!\!$ & $\!\!56.7\!\!$\\
   & $\!\!6\!\!$ & $\!2\!$ & $\!3\!$ & $\!\!1903\!\!$ & $\!\!11.7\!\!$ & $\!\!72.8\!\!$\\
   & $\!\!6\!\!$ & $\!2\!$ & $\!3\!$ & $\!\!1904\!\!$ & $\!\!31.1\!\!$ & $\!\!75.9\!\!$\\
   & $\!\!6\!\!$ & $\!2\!$ & $\!3\!$ & $\!..\!$ & .. & ..\\
   \end{tabular}
}\endgroup
\end{subfigure}\vspace{5pt}\\
\vspace{-9pt}
\end{framed}
\vspace{-8pt}
\caption{Resources from Example \ref{ex:population}. (Left). Primitive fd sets extracted from the given structures $\mathcal S_k(\mathcal E_k,\, \mathcal V_k)$ for $k=1..3$. (Right). \emph{Certain} relations $\boldsymbol H_3\!=\! \{ H_3^1,\; H_3^2 \}$ of hypothesis $\upsilon\!=\!3$ loaded with trial datasets identified by special attribute \textsf{tid}.}
\label{fig:population}
\vspace{-3pt}
\end{figure*}

\subsection{Synthesis `4C'}
\noindent
\textbf{Design by synthesis.} 
The problem of design by synthesis has long been introduced by Bernstein in purely symbolic terms as follows \cite{bernstein1976}: 
given a set $U\!$ of attribute symbols and a set $\Sigma$ of mappings of sets of symbols into symbols (the fd's), find a collection $\boldsymbol R \!=\! \{R_1, R_2, \!\hdots, R_n\}$ (the relations) of subsets of $U$ and, for each $R_i$, a subset of $R_i$ (its designated key) satisfying properties: (P1) each $R_i \in \boldsymbol R$ is in 3NF; (P2) $\boldsymbol R$ completely characterizes $\Sigma$; and (P3) the cardinality $|\boldsymbol R|$ is minimal. 
More generally, the problem of schema design given dependencies considers criteria \cite[Ch. $\!$11]{abiteboul1995}:
\begin{itemize}
\item[P1$^\prime\!$.] $\!\!$($\simeq\!$ P1). $\!$Desirable properties by normal forms;\vspace{-5pt}
\item[P2$^\prime\!$.] $\!\!$($\simeq\!$ P2). $\!$Preservation of dependencies (``meta-data'');\vspace{-5pt}
\item[P3.] $\!$The cardinality $|\boldsymbol R|$ is minimal (minimize joins);\vspace{-5pt}
\item[P4.] $\!$Preservation of data (the lossless join property).
\end{itemize}

There is a well-known trade-off between P1$^\prime$ and P2$^\prime$, since normal forms that ensure less redundant schemas may lose the property of dependency preservation \cite{abiteboul1995}. In fact, P2$^\prime$ is important to prevent the DB from the so-called update anomalies, as the fd's in $\Sigma_k$ are viewed as integrity constraints to their associated relations. 
Hypothesis management applications \cite{goncalves2014}, however, are OLAP-like and have an ETL procedure characterized by batch-, incremental-only updates and large data volumes. 
Thus, it is a design principle of $\Upsilon$-DB to trade P2$^\prime$ for P1$^\prime$, to favor succintness (as less redundancy as possible) over dependency preservation (viz., we stick to BCNF). We do that in a principled way by reasoning over the fd's to ensure the causal ordering is preserved \cite{goncalves2015a}. Also, we shall favor P4 as less joins means faster access to data. 

Overall, for such a notion of `good' design, the hypothesis causal ordering mapped into fd set $\Sigma_k$ needs to be further processed in terms of \emph{acyclic reasoning on its reflexive pseudo-transitive closure}. The demonstrated system relies on an original, efficient algorithm we have designed for that \cite{goncalves2015a}. It returns an fd set $\Sigma_k^\prime$ we motivate and define to be the \emph{folding} $\Sigma_k^\looparrowright\!$ of $\Sigma_k$. Then it applies a variant of Bernstein's synthesis algorithm (say, for certainty `4C') to render, given $\Sigma^\looparrowright\!$, a relational schema $\boldsymbol H_k\!= \bigcup_{i=1}^n H_k^i$. Every schema produced this way provably satisfies desirable properties for hypothesis management \cite{goncalves2015a}.

\begin{myprob} (\textbf{Synthesis `4C'}).
Given fd set $\Sigma_k$, derive an fd set $\Sigma_k^\prime$ (causal ordering processing) to synthesize a relational schema $\bigcup_{i=1}^{n}H_k^i\!$ over it satisfying P1$^\prime$ (BCNF), P3 and striving for P4 while giving up P2$^\prime$.
\label{prob:synthesis4c}
\end{myprob}

For hypothesis $H_3$, e.g., from $\Sigma_3$ we derive its folding $\Sigma_3^\looparrowright\!=\!\{\, \phi \!\to x_0,\, \phi \!\to b,\, \phi \!\to p,\, \phi \!\to y_0,\, \phi \!\to d,\, \phi \!\to r,\; \phi\,t\,\upsilon\,y \!\to x,\; \phi\,t\,\upsilon\,x \!\to y\,\}$. It is input to $\boldsymbol H_3 \!\!:= \textsf{synthesize}(\Sigma_1^\looparrowright\!)$, which is shown in Fig. \!\ref{fig:population}. $\!\!$Every schema produced by this method provably satisfies the properties just mentioned \cite{goncalves2015a}.

Once schema $\boldsymbol H_k$ is synthesized, datasets $\mathcal D_k^\ell$ computed from the hypothesis under alternative trials (input settings) can be loaded into it to accomplish the \textsf{ETL} phase of the design pipeline. Hypothesis management is then enabled for the user, yet up to the capabilities of a traditional relational DB at this stage of the design pipeline. 
In Example \ref{ex:population}, we consider trial datasets for hypothesis $\upsilon\!=\!3$ (viz., the Lotka-Volterra model), which are loaded into the synthesized (certain) schemes in $\boldsymbol H_3$ (see Fig. $\!$\ref{fig:population}). Note that the fd's in $\Sigma_3$ are violated by relations $H_3^1, \,H_3^2$, but we admit a special attribute `trial id' \textsf{tid} into their key constraints for a trivial repair (provisionally, yet at the \textsf{ETL} stage of the pipeline) until uncertainty is introduced in a controlled way by synthesis `4U' (\textsf{U-intro} stage, cf. Fig. $\!$\ref{fig:pipeline}).

\subsection{Synthesis `4U'}
\noindent
There are two sources of uncertainty considered in the $\Upsilon$-DB system: (i) \emph{theoretical} uncertainty, arising from the multiplicity of hypotheses targeted at explaining a same phenomenon; and (ii) \emph{empirical} uncertainty, the multiplicity of trials (alternative parameter settings) under a same hypothesis in view of a same phenomenon. It should be expected, then, that the \textsf{U-intro} procedure is operated by the system in the `global' view of all available hypotheses $k=1..z$.

\textbf{U-relations}. Synthesis `4U' is performed through a data transformation from `certain' relations to `uncertain' U-rela\-tions in the language of probabilistic world-set algebra (p-$\!$WSA). The latter comprises the operations of relational algebra, an operation for computing tuple confidence \textsf{conf}, and the \textsf{repair-key} operation for introducing uncertainty ---  by giving rise to alternative worlds as maximal-subset repairs of an argument key \cite{koch2009}.  
For a prompt illustration, consider query $Q_0$, which we present first as p-WSA formula (\ref{eq:repair-key}), and then in \textsf{MayBMS}'s extended SQL concrete syntax. Note that relation $H_0$ stores (as foreign keys) all hypotheses available in Example \ref{ex:population} and their target phenomena, see Fig. $\!$\ref{fig:maybms}.
\begin{eqnarray}\label{eq:repair-key}
\! Y_0 \,:=\, \pi_{\phi,\upsilon}( \textsf{repair-key}_{\phi @\textsf{Conf}} (H_0)\,).
\end{eqnarray}
\indent\indent\indent
$Q_0$. \textsf{\textbf{create table} $Y_0$ \textbf{as select} $\phi,\,\upsilon$ \textbf{from}\\ \indent\indent\indent\indent\indent(\textbf{repair key} $\phi$ \textbf{in} $H_0$ \textbf{weight by} Conf);}
\vspace{9pt}
\vspace{-3pt}

Query $Q_0$'s result set is materialzed into U-relational table \textsf{Y}$_0$ (Fig. \ref{fig:maybms}). 
U-relations have in their schema a set of pairs $(V_i, D_i)$ of \emph{condition columns} \cite{koch2009} to map each discrete random variable $\textsf{x}_i$ created by the \textsf{repair-key} operation to one of its possible values (e.g., $\textsf{x}_0 \mapsto 1$). The world table $W$, inspired in pc-tables \cite{suciu2011}, 
stores their marginal probabilities.

\begin{figure}[b]
\centering
\scriptsize
\begingroup\setlength{\fboxsep}{2pt}
\colorbox{blue!7}{%
   \begin{tabular}{c|c|c}
  \textsf{H}$_0$ & $\phi$ & $\upsilon\!\!\!$\\
      \hline    
   & $1$ & $1\!\!\!$\\
   & $1$ & $2\!\!\!$\\
   \cline{2-3}
   & $2$ & $1\!\!\!$\\
   & $2$ & $2\!\!\!$\\
   & $2$ & $3\!\!\!$\\
   \end{tabular}
}\endgroup
\hspace{1pt}
\begingroup\setlength{\fboxsep}{2pt}
\colorbox{yellow!15}{%
   \begin{tabular}{c|>{\columncolor[gray]{0.92}}c|c||c}
  $Y_0$ & $\!\!\!V \mapsto D\!\!\!$ & $\phi$ & $\upsilon\!\!\!$\\
      \hline    
   & $\!\!\textsf{x}_0 \mapsto 1\!\!$ & $1$ & $1\!\!\!$\\
   & $\!\!\textsf{x}_0 \mapsto 2\!\!$ & $1$ & $2\!\!\!$\\
   \cline{2-4}
   & $\!\!\textsf{x}_1 \mapsto 1\!\!$ & $2$ & $1\!\!\!$\\
   & $\!\!\textsf{x}_1 \mapsto 2\!\!$ & $2$ & $2\!\!\!$\\
   & $\!\!\textsf{x}_1 \mapsto 3\!\!$ & $2$ & $3\!\!\!$\\
   \end{tabular}
}\endgroup
\hspace{1pt}
\begingroup\setlength{\fboxsep}{2pt}
\colorbox{yellow!15}{%
   \begin{tabular}{c|>{\columncolor[gray]{0.92}}c||c}
  $W$ & $\!\!\!V \mapsto D\!\!\!$ & \textsf{Pr}$\!\!\!$\\
      \hline    
   & $\!\!\textsf{x}_0 \mapsto 1\!\!$ & $.5\!\!\!$\\
   & $\!\!\textsf{x}_0 \mapsto 2\!\!$ & $.5\!\!\!$\\
   \cline{2-3}
   & $\!\!\textsf{x}_1 \mapsto 1\!\!$ & $.33\!\!\!$\\
   & $\!\!\textsf{x}_1 \mapsto 2\!\!$ & $.33\!\!\!$\\
   & $\!\!\textsf{x}_1 \mapsto 3\!\!$ & $.33\!\!\!$\\
   \end{tabular}
}\endgroup
\vspace{-4pt}
\caption{U-relation $Y_0$ rendered by repair-key in the `global' view for all hypotheses and their target phenomena.}
\label{fig:maybms}
\vspace{-8pt}
\end{figure}

The \textsf{U-intro} procedure is based on the fd's but also on the data. It processes the uncertainty of `input' relations (\emph{u-factor\-}\emph{ization}) and then propagates it onto `output' relations (\emph{u-propagation}) \cite{goncalves2015a}. The former is motivated by a basic design principle, which is to define exactly one random variable for each actual uncertainty factor (`u-factor' for short). 
The multiplicity of (competing) hypotheses is itself a standard one, the \emph{theoretical} u-factor (cf. $H_0$ in Fig. $\!$\ref{fig:maybms}). 
The multiplicity of (competing) trials of a hypothesis $k$ for a phenomenon leads to a problem of learning \emph{empirical} u-factors in the local view of each hypothesis $\boldsymbol H_k \subseteq \boldsymbol H$. It is dominated by the problem of fd's discovery in a relation, which has been extensively studied in the literature (e.g., see \cite{huhtala1999}). 

The system then carries out the same kind of reasoning but over a different fd set, formed by merging the primitive fd set, $\Sigma_k$, with the learned (contingent) fd's \cite{goncalves2015a}. It is input, together with $\boldsymbol H_k$ itself, to a different synthesis procedure (now, for uncertainty `4U') to render U-relations. 
\vspace{-3pt}
\begin{myprob} (\textbf{Synthesis `4U'}).
Given a collection of relations $\bigcup_{k=1}^{z}\bigcup_{i=1}^{n}H_k^i$ loaded with alternative trial datasets $\bigcup_{k=1}^{z}\bigcup_{\ell=1}^p \mathcal D_k^\ell$, introduce properly all the uncertainty implicitly present in the database into U-relations $\bigcup_{k=1}^{z}\bigcup_{j=1}^{m}Y_k^j$.
\label{prob:synthesis4u}
\vspace{-8pt}
\end{myprob}
\noindent
Fig. $\!$\ref{fig:u-relations} shows the rendered U-relations for hypothesis $\upsilon\!=\!3$, whose relations are shown in Fig. $\!$\ref{fig:population}. $\!$Note that $\textsf{tid}\!=\!6$ in $H_3^2$ (Fig. $\!$\ref{fig:population}) corresponds now to $\theta = \{\,\textsf{x}_1 \!\mapsto\! 3,\, \textsf{x}_2 \!\mapsto\! 1,$ $\textsf{x}_3 \!\mapsto\! 3,\, \textsf{x}_4 \!\mapsto\! 2 \,\}$, where $\theta$ defines a particular world whose probability is $\textsf{Pr}(\theta) \!\approx\! .055$. This value is computed by the \textsf{conf} aggregate operation based on the marginal probabilities stored in world table $W\!$, as illustrated in the demo (cf. \S\ref{sec:demo}).

\begin{figure}[t]
\centering
\scriptsize
\begingroup\setlength{\fboxsep}{2pt}
\colorbox{yellow!10}{%
   \begin{tabular}{c|>{\columncolor[gray]{0.92}}c||c|c}
  $\!\!$\textsf{Y}$_3^1\!\!$ & $\!\!\!V \!\mapsto\! D\!\!\!$ & $\!\!\!\phi\!\!\!$ & $\!\!\!x_0\!\!\!$\\
      \hline    
   & $\!\!\textsf{x}_2 \!\mapsto\! 1\!\!$ & $\!\!2\!\!$ & $\!\!30\!\!$\\
   \end{tabular}
}\endgroup
\begingroup\setlength{\fboxsep}{2pt}
\colorbox{yellow!10}{%
   \begin{tabular}{c|>{\columncolor[gray]{0.92}}c||c|c}
  $\!\!$\textsf{Y}$_3^2\!\!$ & $\!\!\!V \!\mapsto\! D\!\!\!$ & $\!\!\phi\!\!$ & $\!\!\!b\!\!\!$\\
      \hline    
   & $\!\!\textsf{x}_3 \!\mapsto\! 1\!\!$ & $\!2\!$ & $\!\!.5\!\!\!$\\
   & $\!\!\textsf{x}_3 \!\mapsto\! 2\!\!$ & $\!2\!$ & $\!\!.4\!\!\!$\\
   & $\!\!\textsf{x}_3 \!\mapsto\! 3\!\!$ & $\!2\!$ & $\!\!\!.397\!\!\!\!$\\
   \end{tabular}
}\endgroup
\begingroup\setlength{\fboxsep}{2pt}
\colorbox{yellow!10}{%
   \begin{tabular}{c|>{\columncolor[gray]{0.92}}c||c|c}
  $\!\!$\textsf{Y}$_3^3\!\!$ & $\!\!\!V \!\mapsto\! D\!\!\!$ & $\!\!\phi\!\!$ & $\!\!\!p\!\!\!$\\
      \hline    
   & $\!\!\textsf{x}_4 \!\mapsto\! 1\!\!$ & $\!2\!$ & $\!\!.020\!\!\!\!$\\
   & $\!\!\textsf{x}_4 \!\mapsto\! 2\!\!$ & $\!2\!$ & $\!\!\!.018\!\!\!\!$\\
   \end{tabular}
}\endgroup
\vspace{1pt}\\
\begingroup\setlength{\fboxsep}{3pt}
\colorbox{yellow!10}{%
   \begin{tabular}{c|>{\columncolor[gray]{0.92}}c|>{\columncolor[gray]{0.92}}c|>{\columncolor[gray]{0.92}}c|>{\columncolor[gray]{0.92}}c||c|c|c|c|c}
  $\!\!$\textsf{Y}$_3^4\!\!$ & $\!\!\!V_1 \!\mapsto\! D_1\!\!\!$ & $\!\!\!V_2 \!\mapsto\! D_2\!\!\!$ & $\!\!\!V_3 \!\mapsto\! D_3\!\!\!$ & $\!\!\!V_4 \!\mapsto\! D_4\!\!\!$ & $\!\phi\!$ & $\!\upsilon\!$ & $\! t \!$ & $y$ & $x$\\
      \hline    
   & $\!\!\textsf{x}_1 \mapsto 3\!\!$ & $\!\!\textsf{x}_2 \mapsto 1\!\!$ & $\!\!\textsf{x}_3 \mapsto 1\!\!$ & $\!\!\textsf{x}_4 \mapsto 1\!\!$ & $\!2\!$ & $\!3\!$ & $\!\!1900\!\!$ & $\!\!4\!\!$ & $\!\!30\!\!$\\
   & $\!\!\textsf{x}_1 \mapsto 3\!\!$ & $\!\!\textsf{x}_2 \mapsto 1\!\!$ & $\!\!\textsf{x}_3 \mapsto 1\!\!$ & $\!\!\textsf{x}_4 \mapsto 1\!\!$ & $\!2\!$ & $\!3\!$ & $\!...\!$ & ... & ...\\
   \cline{2-10}
   & $\!\!...\!\!$ & $\!\!...\!\!$ & $\!\!...\!\!$ & $\!\!...\!\!$ & $\!2\!$ & $\!3\!$ & $\!...\!$ & ... & ...\\
   \cline{2-10}
   & $\!\!\textsf{x}_1 \mapsto 3\!\!$ & $\!\!\textsf{x}_2 \mapsto 1\!\!$ & $\!\!\textsf{x}_3 \mapsto 3\!\!$ & $\!\!\textsf{x}_4 \mapsto 2\!\!$ & $\!2\!$ & $\!3\!$ & $\!\!1900\!\!$ & $4$ & $\!\!30\!\!$\\
   & $\!\!\textsf{x}_1 \mapsto 3\!\!$ & $\!\!\textsf{x}_2 \mapsto 1\!\!$ & $\!\!\textsf{x}_3 \mapsto 3\!\!$ & $\!\!\textsf{x}_4 \mapsto 2\!\!$ & $\!2\!$ & $\!3\!$ & $\!\!1901\!\!$ & $\!\!4.12\!\!$ & $\!\!41.5\!\!$\\
   & $\!\!\textsf{x}_1 \mapsto 3\!\!$ & $\!\!\textsf{x}_2 \mapsto 1\!\!$ & $\!\!\textsf{x}_3 \mapsto 3\!\!$ & $\!\!\textsf{x}_4 \mapsto 2\!\!$ & $\!2\!$ & $\!3\!$ & $\!\!1902\!\!$ & $\!\!5.78\!\!$ & $\!\!56.7\!\!$\\
   & $\!\!\textsf{x}_1 \mapsto 3\!\!$ & $\!\!\textsf{x}_2 \mapsto 1\!\!$ & $\!\!\textsf{x}_3 \mapsto 3\!\!$ & $\!\!\textsf{x}_4 \mapsto 2\!\!$ & $\!2\!$ & $\!3\!$ & $\!\!1903\!\!$ & $\!\!11.7\!\!$ & $\!\!72.8\!\!$\\
   & $\!\!\textsf{x}_1 \mapsto 3\!\!$ & $\!\!\textsf{x}_2 \mapsto 1\!\!$ & $\!\!\textsf{x}_3 \mapsto 3\!\!$ & $\!\!\textsf{x}_4 \mapsto 2\!\!$ & $\!2\!$ & $\!3\!$ & $\!\!1904\!\!$ & $\!\!31.1\!\!$ & $\!\!75.9\!\!$\\
   & $\!\!\textsf{x}_1 \mapsto 3\!\!$ & $\!\!\textsf{x}_2 \mapsto 1\!\!$ & $\!\!\textsf{x}_3 \mapsto 3\!\!$ & $\!\!\textsf{x}_4 \mapsto 2\!\!$ & $\!2\!$ & $\!3\!$ & $\!..\!$ & .. & ..\\
   \end{tabular}
}\endgroup
\vspace{-4pt}
\caption{U-relations rendered for hypothesis $\upsilon=3$.}
\label{fig:u-relations}
\vspace{-15pt}
\end{figure}

\subsection{Conditioning}
Possible-worlds semantics can be seen as a generalization of \emph{data cleaning}. In the context of p-DB's \cite{suciu2011}, data cleaning does not have to be (error-prone) one shot \cite{beskales2009}. Rather, it can be carried out gradually, viz., by keeping all mutually inconsistent tuples under a probability distribution (ibid.) that can be updated in face of evidence until the probabilities of some tuples eventually tend to zero to be eliminated. This leads us to Remark \ref{rmk:method}, motivating the hypothesis analytics feature of the demonstrated system.

\begin{myremark}
\label{rmk:method}
Consider U-relational table $Y_0$ (Fig. \ref{fig:maybms}). Note that it abstracts the goal of a data-driven hypothesis evaluation study as the repair of each $\phi$ as a key. The \emph{\mbox{$\Upsilon$-DB}} system is designed for enabling users to develop their research with sup\-port of query and update capabilities such that they can manage and rate/rank their hypotheses $\upsilon$ w.r.t.$\!$ each $\phi$, until the relationship $r(\phi,\, \upsilon)$ is eventually repaired to be a function $f\!: \Phi \to \Upsilon$ from each phenomenon $\phi $ to its best explanation $\upsilon$. $\Box$
\end{myremark}

\textbf{Bayesian inference}. 
\noindent
As suggested by Remark \ref{rmk:method}, the prior probability distribution assigned via \mbox{\textsf{repair key}} is to be eventually conditioned on observed data. 
The \mbox{$\Upsilon$-DB} system enables the user to carry out Bayesian inference steps that update at each step a prior to a posterior. Our primary use case is in computational science (cf. \S\ref{sec:demo}), where we have \emph{discrete} random variables mapped to the possible values of (numerical) prediction attributes whose domain are \emph{continuous} (double precision). Therefore Bayesian inference is applied for normal mean with a discrete prior \cite{bolstad2007}.

The procedure uses normal density function (\ref{eq:density}) with standard deviation \mbox{$\sigma$} to compute the likelihood $f(y \,|\, \mu_k)$ for each competing prediction (trial) $\mu_k$ given observation $y$. But in the general case we actually have 
a sample of independent observed values $y_1,\,...,\,y_n$ (e.g., Lynx population in Hudson's Bay over the years). Then, the likelihood $f(y_1,\,...,\,y_n \,|\, \mu_k)$ for each competing trial $\mu_k$, is computed as a product $\textstyle\prod_{j=1}^n f(y_j \,|\, \mu_{kj})$ of the single likelihoods $f(y_j \,|\, \mu_{kj})$ \cite{bolstad2007}. Bayes' rule is then settled by (\ref{eq:sample-bayes}) to compute the posterior $p(\mu_k \,|\, y_1,\,...,\,y_n)$ given prior $p(\mu_k)$. 

\vspace{-5pt}
\begin{eqnarray}
f(y \,|\, \mu_k) \!\!&=&\!\! \frac{1}{\sqrt{2 \pi \sigma^2}}\, e^{-\frac{1}{2\sigma^2}(y-\mu_k)^2}\label{eq:density}
\end{eqnarray}
\begin{eqnarray}
p(\mu_k \,|\, y_1,\,\hdots,\,y_n) \!&=&\! \frac{\textstyle\prod_{j=1}^n f(y_j \,|\, \mu_{kj})\;p(\mu_k)}{\displaystyle\sum_{i=1}^m \displaystyle\prod_{j=1}^n f(y_j \,|\, \mu_{ij})\;p(\mu_i)}\label{eq:sample-bayes}
\end{eqnarray}

\setcounter{myex}{0}
\begin{myex}\label{ex:bayes}$\!$(continued).$\!$ 
Consider observational data collected from Hudson's Bay from 1900 to 1920 for the Lynx and Hare populations at each year \cite{elton1942}. Suppose that the user queries the hypothesis predictions available in the system (viz., by mapping $t \mapsto `Year'$, and $x \mapsto `Lynx'$) at, say, year 1904. For rating/ranking the predictions, their (prior) probabilies are conditioned on the available observations and listed (posteriors) for supporting an informed decision about the underlying hypotheses, see Fig. $\!$\ref{fig:analytics}. Here, the probabilities are computed taking into account all sample points from year 1900 to 1920, not only the slected year of $\!$1904. $\Box$
\end{myex}
%
\begin{figure}[t]\scriptsize
\centering
\begingroup\setlength{\fboxsep}{3pt}
\colorbox{yellow!15}{%
   \begin{tabular}{c|c|c|c|c|c|>{\columncolor[gray]{0.92}}c||>{\columncolor[gray]{0.92}}c}
  \textsf{Y[$s$]} & $\phi$ & $\upsilon$ & \textsf{tid} & \textsf{Year} & \textsf{Lynx} & \textsf{Prior} & \textsf{Posterior}\\
      \hline    
   & $2$ & $1$ & $1$ & $1904$ & $16.49$ & $.167$ & $.047$\\
   & $2$ & $1$ & $2$ & $1904$ & $18.22$ & $.167$ & $.000$\\
\cline{2-6}
   & $2$ & $2$ & $1$ & $1904$ & $79.81$ & $.167$ & $.013$\\
   & $2$ & $2$ & $2$ & $1904$ & $77.82$ & $.167$ & $.017$\\
\cline{2-6}
   & $2$ & $3$ & $1$ & $1904$ & $89.59$ & $.055$ & $.131$\\
   & $2$ & $3$ & $2$ & $1904$ & $65.06$ & $.055$ & $.184$\\
   & $2$ & $3$ & $3$ & $1904$ & $90.08$ & $.055$ & $.124$\\
   & $2$ & $3$ & $4$ & $1904$ & $77.46$ & $.055$ & $.176$\\
   & $2$ & $3$ & $5$ & $1904$ & $88.32$ & $.055$ & $.127$\\
   & $2$ & $3$ & $6$ & $1904$ & $75.92$ & $.055$ & $.180$\\
   \end{tabular}
}\endgroup\vspace{-3pt}  
\caption{$\Upsilon$-DB query for data-driven hypothesis analytics.}\label{fig:analytics}
\vspace{-8pt}
\end{figure}
%
The aggregate posterior of each hypothesis as an explanation for a given phenomenon is given by the sum of the posteriores of its alternative predictions, e.g., \textsf{Pr}$_{\upsilon \,\mapsto\, 2} \!\approx\! 030$.

\begin{myprob} (\textbf{Conditioning}).
Given a prior probability distribution assigned to competing predictions for a target phenomenon and stored as marginal probabilities in the world table $W\!$, apply Bayesian inference and induce effects of posteriors back to table $W\!$.
\label{prob:conditioning}
\end{myprob}

Note that this is an applied \emph{Bayesian inference} problem that translates into a \emph{p-$\!$DB update} one to induce effects of posteriors back to table $W\!$. In this first prototype of \mbox{$\Upsilon$-DB} we accomplish it by performing Bayesian inference at application level and then applying p-WSA's \textsf{update} (a variant of SQL's \textsf{update}) into \textsf{MayBMS}. It is a topic of future work to design a dedicated algebraic operation for conditioning.

\vspace{-2pt}
\section{Related Work}\label{sec:related-work}
\noindent
\textbf{Models and data}. 
Haas et al. \cite{haas2011} propose a long-term \emph{models-and-data} research program to extend data management technology for `prescriptive' analytics. This is (sic.) to identify optimal business, policy, investment, and engineering decisions in the face of uncertainty. Such analytics shall rest on deep `predictive' analytics that go beyond mere statistical forecasting and are imbued with an understanding of the fundamental mechanisms that govern a system's behavior, allowing what-if analyses \cite{haas2011}. They discusss strategies to extend query engines for model execution within a \mbox{(p-)DB}. Along these lines, query optimization is understood as a more general problem with connections to algebraic solvers.
Our framework in turn essentially comprises an abstraction of \emph{hypotheses as data} \cite{goncalves2014}. It can be understood in comparison as putting models strictly into a (flattened) data perspective. Therefore, and as evidenced by this prototype demonstration, it is directly applicable by building upon recent work on p-DBs \cite{suciu2011}.

\textbf{Design by synthesis}. 
Classical design by synthesis \cite{bernstein1976} was once criticized due to its too strong `uniqueness' of fd's assumption, 
as it reduces the problem of design to symbolic reasoning on fd's arguably neglecting semantic issues. Hypothesis management, however, sets a use case where design by synthesis is clearly feasible, as it translates seamlessly to data dependencies the reduction made by the user herself into a tentative formal model for the studied phenomenon. In fact, in such a class of applications synthesis methods may be as fruitful for p-DB design as they are for the design of graphical models (cf. \cite{darwiche2010}).

\textbf{Causality in DBs}. 
Our encoding of equations into fd's (constraints/correlations) captures the causal chain from exogenous (input) to endogenous (output) tuples. Fd's are rich, stronger information that can be exploited in reasoning about causality in a DB for the sake of explanation and sensitivity analysis. 
$\!$In comparison with \mbox{Meliou et. al \cite{meliou2010}}, 
we are processing causality at schema level. To our knowledge this is the first work to address causal reasoning in the presence of constraints (viz., fd's). 

\textbf{Conditioning}. 
The topic of conditioning a p-DB has been firstly addressed by Koch and Olteanu motivated by data cleaning applications \cite{koch2008}. They have introduced the \texttt{assert} operation to implement, as in AI, a kind of knowledge compilation, viz., world elimination in face of constraints (e.g., FDs). For hypothesis management in $\Upsilon$-DB, nonetheless, there is a need to apply \emph{Bayes' conditioning} by asserting observed data (not constraints).

\textbf{Hypothesis encoding}. 
Finally, our framework is comparable with Bioinformatics' initiatives that address hypothesis encoding into the RDF data model \cite{soldatova2011}.  
We point out the Robot Scientist, HyBrow and SWAN (cf. \cite{soldatova2011}), all of which consist in some ad-hoc RDF encoding of sequence and genome analysis hypotheses under varying levels of structure (viz., from `gene G has function A' statements to free text). Our framework in turn consists in the U-relational encoding of hypotheses from mathematical equations, which is (to our knowledge) the first work on hypothesis relational encoding.

\vspace{-3pt}
\section{Applicability}\label{sec:applicability}
\noindent
\textbf{Realistic assumptions}. The core assumption of the $\Upsilon$-DB system is that the deterministic hypotheses, given in a \textsf{MathML}-compliant XML file, are structured (in equations) such that they are encodable into a SEM that must be valid (i.e., $|\mathcal E| \!=\! |\mathcal V|$). Also, as a semantic assumption which is standard in scientific modeling, we consider a one-to-one correspondence between real-world entities and variable/at\-tribute symbols within a structure, and that all of them must appear in some of its equations/fd's. 
For most science use cases (if not all) involving deterministic models, such assumptions are quite reasonable. It is a topic of future work to explore business use cases as well, and qualitative hypotheses as discussed elsewhere \cite{goncalves2015a}. 

\textbf{Hypothesis learning}. The (user) method for hypothesis generation is irrelevant to our framework, as long as the resulting hypotheses are encodable into a valid SEM. Thus, promising use cases may arise through the incorporation of machine learning methods into our framework to scale up the formation/extraction of hypotheses and evaluate them under the querying capabilities of a p-DB. Consider, e.g., learning the equations from \textsf{Eureqa}.\footnote{\url{http://creativemachines.cornell.edu/Eureqa}.}

$\!$\textbf{Model repositories}. $\!$Recent initiatives have been fostering large-scale model integration, sharing and reproducibility in the computational sciences.\footnote{\url{http://senselab.med.yale.edu/modeldb/}.}\textsuperscript{,}\footnote{\url{http://physiome.org/}.} They are growing reasonably fast on the web, (i) promoting some \textsf{MathML}-based standard for model specification, but (ii) with limited integrity and lack of support for rating/ranking competing models. For those two reasons, they provide a 
strong use case for $\Upsilon$-DB. The Physiome project, 
e.g., is planned to integrate very large deterministic models of human physiology \cite{hunter2003}. A fairly simple model of the human cardiovascular system has 630+ variables (or equations, as $\mathcal |\mathcal E| \!=\! \mathcal |\mathcal V|$), motivating then data-driven hypothesis management and analytics.

\begin{figure*}[t]
\centering
\begin{subfigure}[t]{0.33\textwidth}
\advance\leftskip0.9cm
\includegraphics[width=.69\textwidth]{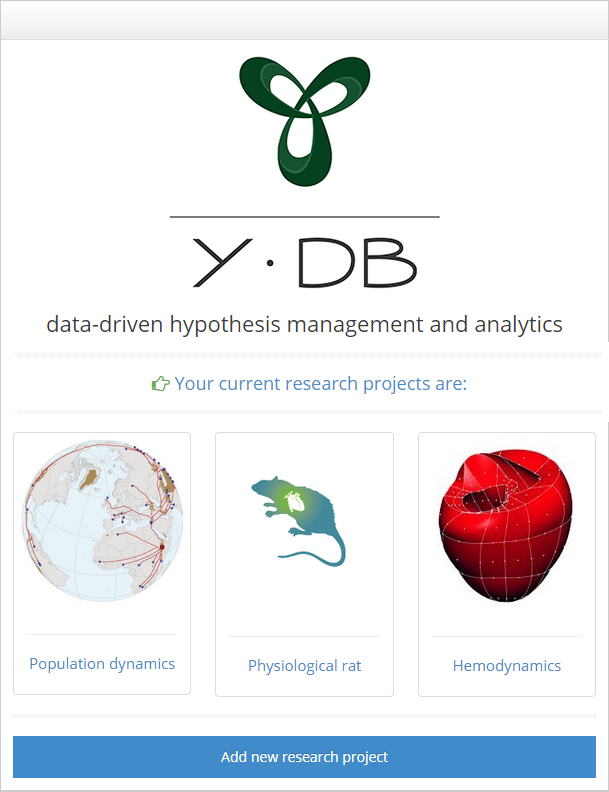}
\vspace{-1pt}
\caption{Research dashboard after login.}
\label{fig:logo}
\end{subfigure}
\hspace{-10pt}
\begin{subfigure}[t]{0.33\textwidth}
\advance\leftskip0.75cm
\includegraphics[width=.78\textwidth]{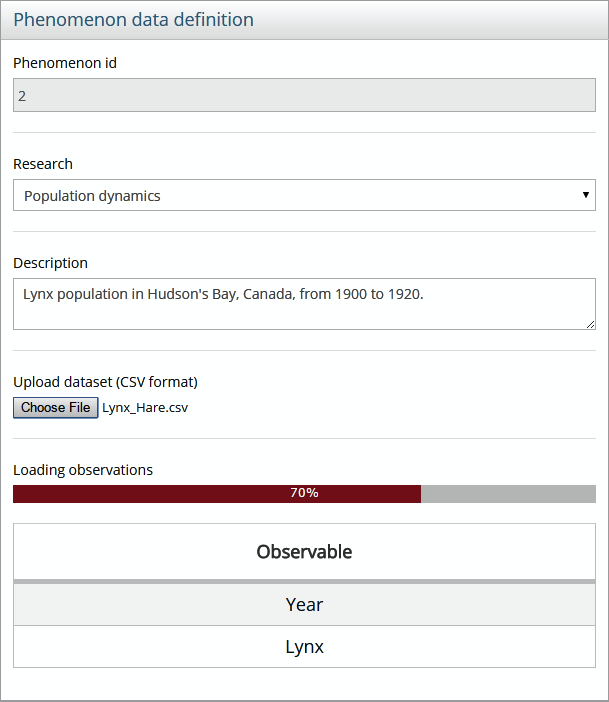}
\caption{Phenomenon data definition.}
\label{fig:etl-phenomenon}
\end{subfigure}
\begin{subfigure}[t]{0.33\textwidth}
\advance\leftskip0.9cm
\includegraphics[width=.7415\textwidth]{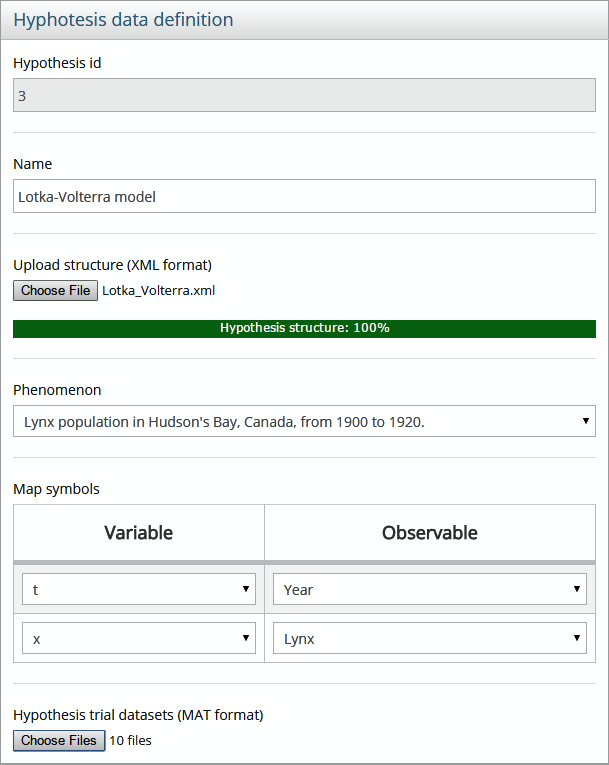}
\caption{Hypothesis data definition.}
\label{fig:etl-hypothesis}
\end{subfigure}\vspace{4pt}\\
\begin{subfigure}[t]{0.33\textwidth}
\advance\leftskip1cm
\includegraphics[width=.705\textwidth]{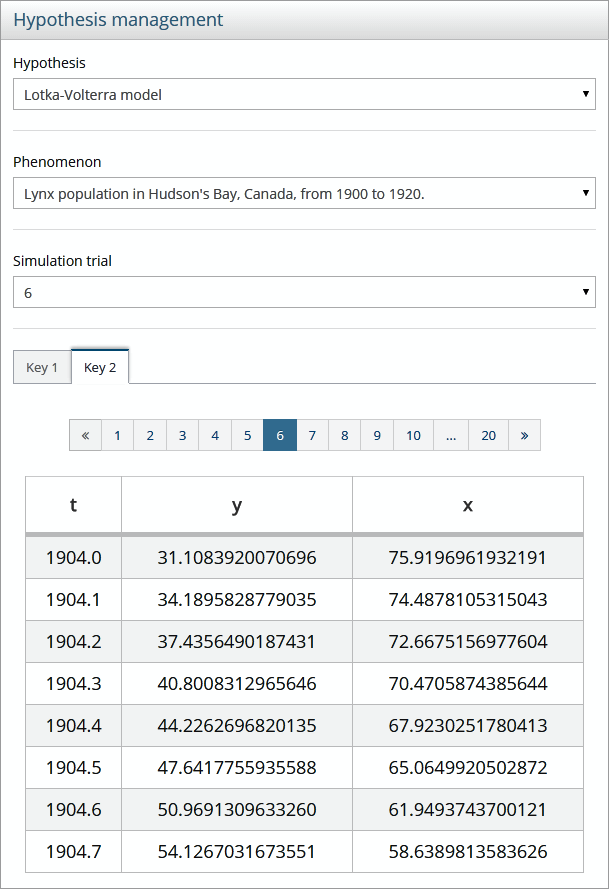}
\caption{Hypothesis management.}
\label{fig:management}
\end{subfigure}
\begin{subfigure}[t]{0.33\textwidth}
\advance\leftskip0.6cm
\includegraphics[width=.81\textwidth]{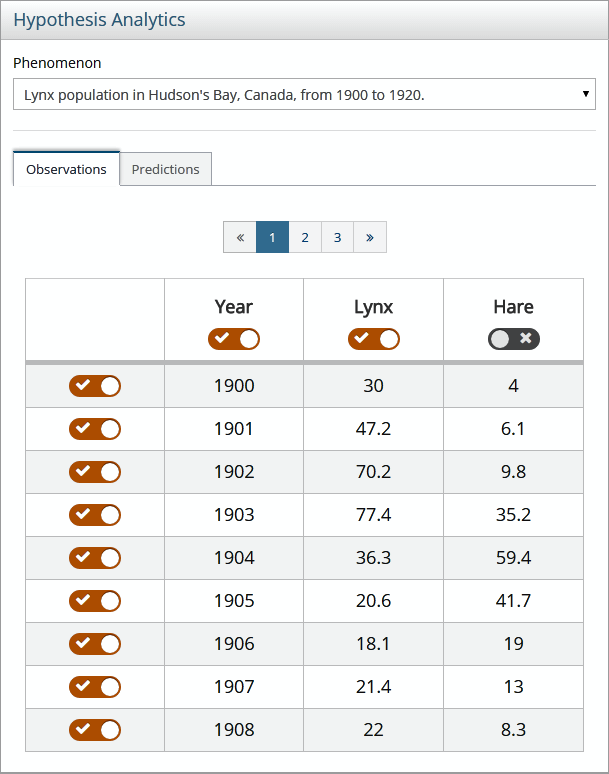}
\caption{Analytics: selected observations tab.}
\label{fig:analytics1}
\end{subfigure}
\hspace{-5pt}
\begin{subfigure}[t]{0.33\textwidth}
\advance\leftskip0.75cm
\includegraphics[width=.775\textwidth]{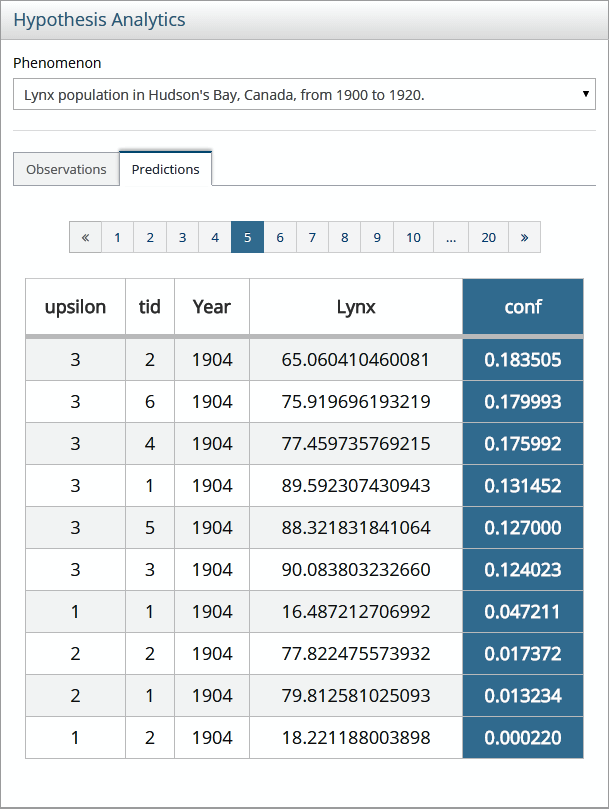}
\caption{Analytics: ranked predictions tab.}
\label{fig:analytics2}
\end{subfigure}
\vspace{-5pt}
\caption{Screenshots of this first prototype of the $\Upsilon$-DB system.}
\label{fig:demo}
\vspace{-9pt}
\end{figure*}

\vspace{-4pt}
\section{Demonstration}\label{sec:demo}
\noindent
This first prototype of $\Upsilon$-DB is implemented as a Java web application, with the pipeline component in the server side on top of \textsf{MayBMS} (a backend extension of \textsf{PostgreSQL}). 

In this demonstration we go through the whole pipeline (Fig. $\!$\ref{fig:pipeline}), exploring our use case scenarios. In addition to the population dynamics scenario presented throughout this text, the demonstration will include scenarios of computational hemodynamics research, where the hypotheses are extracted from the Physiome model repository, and a scenario extracted from the Virtual Physiological Rat project.\footnote{\url{http://virtualrat.org/}.}

The demonstration will unfold in three phases. In the first phase, we show the ETL process to give a sense of what the user has to do in terms of simple phenomena description, hypothesis naming and file upload to get her phenomena and hypotheses available in the system to be managed as data. In the second phase, we will reproduce some typical queries of hypothesis management (e.g., what hypotheses are available for a selected phenomenon, or listing all the predictions under some selectivity criteria, e.g., predicted values of, say, arterial pressure in systole periods). In the third phase, we enter the hypothesis analytics module. The user will choose a phenomenon for a hypothesis evaluation study, and the system will list all the predictions with their probabilities under some selectivity criteria (e.g., population at year 1920). The predictions are ranked according to their probabilities, which are conditioned on the observational data available for the chosen phenomenon.

Fig. $\!$\ref{fig:demo} shows screenshots of the system. Fig. $\!$\ref{fig:demo}(a) shows the research projects currently available for a user. Figs. $\!$\ref{fig:demo}(b, c) show the ETL interfaces for phenomenon and hypothesis data definition (by synthesis), and then the insertion of hypothesis trial datasets, i.e., explanations of a hypothesis towards a target phenomenon. Fig. $\!$\ref{fig:demo}(d) shows the interface for basic hypothesis management by listing the predictions of a given simulation trial. Figs. $\!$\ref{fig:demo}(e, f) show two tabs of the hypothesis analytics module, viz., selection of observations and then viewing the corresponding alternative predictions ranked by their conditioned probabilities.

\vspace{-5pt}
\section{Acknowledgments}
\begin{footnotesize}
\vspace{-3pt}
\noindent
This work has been supported by the Brazilian funding agencies CNPq (grants $\!$n$^o\!$ 141838/2011-6, 309494/2012-5) and FAPERJ (grants INCT-MACC E-26/170.030/2008, `Nota $\!$10' $\!$E-26/100.286/2013). $\!$We thank IBM for a Ph.D. Fellowship 2013-2014.
\end{footnotesize}
\bibliographystyle{abbrv}
\bibliography{sigmod15}

\begin{thebibliography}{10}

\bibitem{abiteboul1995}
S.~Abiteboul, R.~Hull, and V.~Vianu.
\newblock {\em Foundations of Databases}.
\newblock Addison-Wesley, 1995.

\bibitem{bernstein1976}
P.~Bernstein.
\newblock Synthesizing third normal form relations from functional
  dependencies.
\newblock {\em ACM Trans. on Database Systems}, 1(4):277--98, 1976.

\bibitem{beskales2009}
G.~Beskales, M.~A. Soliman, I.~F. Ilyas, and S.~Ben-David.
\newblock Modeling and querying possible repairs in duplicate detection.
\newblock {\em PVLDB}, 2(1):598--609, 2009.

\bibitem{bolstad2007}
W.~M. Bolstad.
\newblock {\em Introduction to Bayesian Statistics}.
\newblock Wiley-Interscience, 2nd edition, 2007.

\bibitem{darwiche2010}
A.~Darwiche.
\newblock $\!${B}ayesian networks.
\newblock {\em $\!$Comm. $\!$ACM}, $\!$53(12):80--90, $\!$2010.

\bibitem{druzdzel2008}
D.~Dash and M.~J. Druzdzel.
\newblock A note on the correctness of the causal ordering algorithm.
\newblock {\em Artif. Intell.}, 172(15):1800--8, 2008.

\bibitem{elton1942}
C.~Elton and M.~Nicholson.
\newblock The ten-year cycle in numbers of the lynx in {C}anada.
\newblock {\em Journal of Animal Ecology}, 11(2):215--44, 1942.

\bibitem{goncalves2014}
B.~Gon\c{c}alves and F.~$\!$Porto.
\newblock {$\Upsilon$-DB}: {M}anaging scientific hypotheses as uncertain data.
\newblock {\em PVLDB}, 7(11):959--62, 2014.

\bibitem{goncalves2015a}
B.~Gon\c{c}alves and F.~Porto.
\newblock Design-theoretic encoding of deterministic hypotheses as constraints
  and correlations in {U}-relational databases.
\newblock In {\em ACM PODS \emph{(UNDER REVIEW,
  \href{http://arxiv.org/abs/1411.5196}{ CoRR abs/1411.5196})}}, 2015.

\bibitem{haas2011}
P.~Haas et~al.
\newblock $\!${Data} is dead... without what-if models.
\newblock {\em PVLDB}, 4(12):1486--9, 2011.

\bibitem{huhtala1999}
Y.~Huhtala et~al.
\newblock {TANE}: An efficient algorithm for discovering functional and
  approximate dependencies.
\newblock {\em Comput. J.}, 42(2):100--11, 1999.

\bibitem{hunter2003}
P.~J. Hunter and T.~K. Borg.
\newblock Integration from proteins to organs: the {P}hysiome {P}roject.
\newblock {\em Nat Rev Mol Cell Biol.}, 4(3):237--43, 2003.

\bibitem{koch2009}
C.~Koch.
\newblock {\em May{BMS}: {A} system for managing large uncertain and
  probabilistic databases}.
\newblock In C. Aggarwal (ed.), Managing and Mining Uncertain Data, Chapter 6.
  Springer-Verlag, 2009.

\bibitem{koch2008}
C.~Koch and D.~Olteanu.
\newblock Conditioning probabilistic databases.
\newblock {\em PVLDB}, 1(1):313--25, 2008.

\bibitem{meliou2010}
A.~Meliou et~al.
\newblock Causality in databases.
\newblock {\em {IEEE} Data Eng. Bull.}, 33(3):59--67, 2010.

\bibitem{simon1953}
H.~Simon.
\newblock {\em Causal ordering and identifiability}.
\newblock In Hood \& Koopmans (eds.), Studies in Eco\-nometric Methods, Chapter
  3, John Wiley \& Sons, 1953.

\bibitem{simon1966}
H.~Simon and N.~Rescher.
\newblock Cause and counterfactual.
\newblock {\em Philosophy of Science}, 33(4):323--40, 1966.

\bibitem{soldatova2011}
L.~Soldatova and A.~Rzhetsky.
\newblock Representation of research hypotheses.
\newblock {\em J Biomed Sem}, 2(S2), 2011.

\bibitem{suciu2011}
D.~Suciu, D.~Olteanu, C.~R\'e, and C.~Koch.
\newblock {\em $\!$Probabilistic$\!$ Databases}.
\newblock Morgan \& Claypool Publishers, 2011.

\end{thebibliography}

\end{document}